# Experimental investigation of the breakdown of the Onsager-Casimir relations

C.A. Marlow[a,*] R.P. Taylor[a], M. S. Fairbanks[a], I. Shorubalko[b], and H. Linke[a,]

[a] *Materials Science Institute, Physics Department, University of Oregon, Eugene OR 97403-1274, USA*
[b] *Institute of Solid State Physics, University of Latvia, Kengarga 8, LV-1063, Riga, Latvia*

We use magnetoconductance fluctuation measurements of phase-coherent semiconductor billiards to quantify the contributions to the nonlinear electric conductance that are asymmetric under reversal of magnetic field. We experimentally determine that the average asymmetric contribution is linear in magnetic field (for magnetic flux much larger than one flux quantum) and that its magnitude depends on billiard geometry. In addition, we find an unexpected asymmetry in the power spectrum characteristics of the magnetoconductance with respect to reversal of magnetic field and bias voltage.

Electron transport in linear response is characterized by a high degree of symmetry with respect to the direction of an applied magnetic field, $B$, as described by the Onsager-Casimir relations $\sigma_{\alpha\beta}(B) = \sigma_{\beta\alpha}(-B)$ in terms of the local conductivity tensor [1]. For the case of two-terminal mesoscopic conductors, this corresponds to the reciprocity theorem, which predicts the conductance to be an even function of $B$ in linear response [2]. At zero magnetic field, it is well known that the absence of spatial symmetry leads to nonlinear terms in the conductance that are asymmetric in $V$ [3]. Only very recently has a distinct nonlinear term attracted attention that leads to an asymmetry of the conductance in the presence of both finite B and V [4,5]. The contribution of electron-electron interaction to this asymmetry was calculated to lowest order for disordered [6] and ballistic [7] phase-coherent conductors for a magnetic flux smaller than one flux quantum $\Phi_o$ = h/e, and for bias voltages, $V$ smaller than the characteristic energy scale of the system (typically μeV in semiconductor billiards).

Here we quantitatively investigate conductance asymmetries in a new regime, namely for magnetic $B$ fields much greater than one flux quantum threaded through the device area, $A$, and for significant bias voltages (on order of mV) using billiards of varying geometry. Using measurements of magnetoconductance fluctuations (MCF) we quantify, for the first time, the lowest nonlinear conductance term that leads to the breakdown of the Onsager-Casimir relations in this regime, and show that its magnitude is related to billiard geometry. In addition we have discovered that not only individual features of the MCF, but, unexpectedly, the characteristics of the MCF power spectrum are asymmetric in $B$ and $V$, and that they depend on device asymmetry.

The nonlinear differential conductance, $g$ $(B, V)$ = dI/dV, can be approximated by an expansion in $V$,

$$g(B,V) \equiv \frac{dI(B,V)}{dV} = g_0(B) + g_1(B)V + g_2(B)V^2 + g_3(B)V^3 ... \quad (1)$$

where $g_0$ $(B)$ is the conductance in linear response. Each coefficient, $g_i$ $(B)$, is the sum of an odd, $g_i^{odd}$ $(B)$, and an even, $g_i^{even}$ $(B)$, function of $B$. The odd contributions to the nonlinear conductance can be isolated by computing the value $\delta^{odd} \equiv g(+B,+V) - g(-B,+V) - g(+B,-V) + g(-B,-V)$ in combination with Eq. (1) [8], which to leading order in $V$ gives,

$$\delta^{odd} = 2 \sum_{i=odd}(g_i(+B) - g_i(-B))V^i = 4g_1^{odd}(B)V + ... \quad (2)$$

In a similar way, the even contributions can be isolated by the sum $\delta^{even} \equiv g(+B,+V) + g(-B,+V) - g(+B,-V) - g(-B,-V)$ which when combined with Eq. (1) is,

$$\delta^{even} = 2 \sum_{i=odd}(g_i(+B) + g_i(-B))V^i = 4g_1^{even}(B)V + ... \quad (3)$$

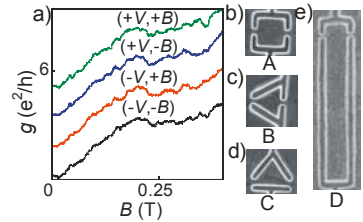

FIG. 1. (a) Magnetoconductance fluctuations measured on device C with an applied bias |$V$|= 1.44 mV for all four combinations of the sign of $V$ and direction of $B$. MCF are offset for clarity. (b) – (e) SEM images of all devices used in this study.

Equations (2) and (3) can then be used to extract the odd and even components of $g_1(B)$ in Eq. (1) from the experimental data [9].

Conductance measurements were made on two-terminal semiconductor billiards. Billiards were defined using electron beam lithography and deep wet-etching of the 2-dimensional electron gas (2DEG) formed in a GaInAs/InP heterostructure (Fermi energy = 35 meV) [10]. This system was chosen for its high shape fidelity in billiard definition [5,10] and because previous studies have shown it exhibits negligible circuit-induced asymmetry ('self-gating'), permitting careful control of device and conductance symmetry [4,5]. Device dimensions were made to be smaller than the electron mean free path such that transport within the billiard was ballistic (see Table 1). Device geometries were designed to have either left-right (LR) or up-down (UD) asymmetry. In the case of LR asymmetry, device symmetry is broken in the direction parallel to the current flow (device B in (Fig. 1(c)) and the device potential is asymmetric under reversal of $V$. For UD asymmetry, symmetry is broken in the direction perpendicular to the current flow (devices C and D in Fig. 1(d) and 1(e)) and the potential is asymmetric with reversal of $B$. Device A (Fig. 1(b)) was included as a control since it nominally contains neither LR nor UD asymmetry. Measurements were made with a $B$ field applied perpendicular to the 2DEG at a temperature $T$ = 240mK with several propagating channels. Transport was phase coherent (see Table 1), leading to magnetoconductance fluctuations (MCF) arising from quantum interference effects. MCF are an excellent tool for probing transport symmetries in the nonlinear regime [5] since they are extremely sensitive to electron scattering configurations [11] and have a high degree of reproducibility. Two-terminal magnetoconductance measurements were made in four-point geometry using lock-in techniques. A tunable dc bias, $V$, was added to a small ac signal ($V_{ac} < kT/e \approx 20\mu V$) and measurements were taken for a variety of billiard geometries for both directions of $B$ at several $\pm V$ and over the range of $B$ = 0 to 0.5 T. Voltages were chosen such that $eV >> E_s$, where $E_s$ is the mean energy level spacing in the billiard, and the $B$ range extending to $B >> \Phi_o/A$.

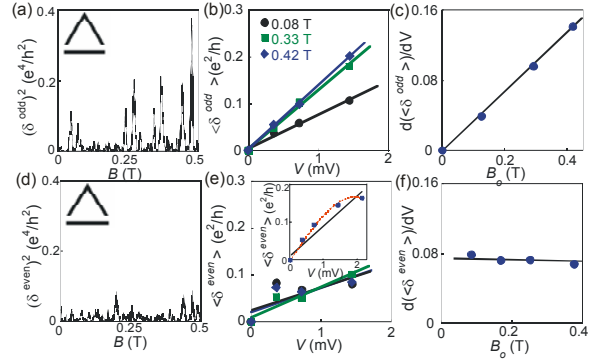

FIG. 2. Asymmetry analysis of MCF for device C. (a) the square of the nonlinear contribution odd in $B$ field, $(\delta^{odd})^2$, shown as a function of $B$, (b) the rms of $\delta^{odd}$, $<\delta^{odd}(B_o)>$, plotted as a function of $V$. For clarity only select $B_o$ values are shown. (c) The slope $d<\delta^{odd}(B_o)>/dV$ determined from the plot in (b) with $B_o$, (d) the square of the nonlinear contribution even in $B$ field, $(\delta^{even})^2$, shown as a function of $B$, (e) $<\delta^{even}(B_o)>$ plotted as a function of $V$, for the same $B_o$ values shown in (b). Inset in (e) the data for $<\delta^{even}(B_o)>$ is replotted this time averaging $<\delta^{even}(B_o)>$ over all $B_o$ as a function of $V$. Any deviations from the linear fit (bold line) are captured by a fit of the form $\gamma V + \beta V^3$ (dashed line) that includes the next highest order term in Eq. (3). (f) the slope $d<\delta^{even}(B_o)>/dV$, determined from the plot in (e) with $B_o$.

First, we determine the functional dependence of the asymmetric nonlinear contributions on $B$. Figure 1(a) shows MCF measured in the four different configurations of $\pm V$ and $\pm B$ for device C at $V$ = 1.44 mV. The MCF data are summed according to Eqs. (2) and (3) to determine $\delta^{odd}$ and $\delta^{even}$, respectively, as a function of $B$. In Fig. 2(a), we plot $(\delta^{odd})^2$ and observe an overall increase with $B$, whereas the even part, $(\delta^{even})^2$, on average does not change significantly with $B$ (Fig. 2(d)). Due to their origin in MCF, $\delta^{odd}$ and $\delta^{even}$ fluctuate with $B$ on the scale of the correlation field $B_c \approx \Phi_o/A \sim 4$ mT [11]. To determine how the asymmetries change across the entire range of $B$, we calculate $<\delta^{odd}(B_o)>$, defined as the rms of $\delta^{odd}(B)$ within a window $[B_o-\Delta B/2; B_o+\Delta B/2]$ centered at $B_o$ where $\Delta B = 0.17$ T $\sim 4B_c$.

| Device (asymmetry) | Area ($\mu m^2$) | $l_\mu$ ($\mu m$) | $l_\phi$ ($\mu m$) $V=0$ | $l_\phi$ ($\mu m$) $V \approx 1$ mV | Ave rms $V=0$ ($e^2/h$) | $<g_0^{odd}>$ ($e^2/h$) | $<g_1^{odd}>$ ($e^2/h$)/m V·T | $<g_1^{even}>$ ($e^2/h$)/mV |
|---|---|---|---|---|---|---|---|---|
| A (-) | 0.72 | 6.2 | 19 | 13 | 0.1 | 0.012 | 0.008 | 0.013 |
| B (LR) | 0.49 | 6.1 | 13 | 11 | 0.1 | 0.012 | 0.015 | 0.035 |
| C (UD) | 0.50 | 6.1 | 16 | 15 | 0.1 | 0.014 | 0.044 | 0.016 |
| D (UD) | 3.5 | 4.0 | 25 | 24 | 0.04 | 0.010 | 0.026 | 0.011 |

Table 1. Experimental details for all devices used in this study: billiard area, $A$ determined from depletion measurements, the electron mean free path, $l_\mu$ determined from electron mobility measurements, the phase-breaking length, $l_\phi$ at $T$ =240 mK for $V$ = 0 and $V$ = 1 mV determined using a correlation field analysis [18], the average rms of the fluctuations in linear response, the noise level $<g_0^{odd}>$ and the calculated nonlinear coefficients $<g_1^{odd}>$ and $<g_1^{even}>$.

We find that $<\delta^{odd}(B_o)>$ increases linearly as a function of $V$ (see Fig. 2(b)), consistent with Eq. (1), confirming that the expansion holds for the dc-bias range used. The rms value of $g_1^{odd}(B)$, $<g_1^{odd}(B)>$, can be determined by plotting the slopes, $d(<\delta^{odd}(B_o)>)/dV$, of the linear fits in Fig. 2(b) as a function of $B_o$ (Fig. 2(c)). A linear dependence of $d(<\delta^{odd}(B_o)>)/dV$ with $B_o$ is observed, indicating $<g_1^{odd}(B)>$ is of the form $<g_1^{odd}(B)> = <g_1^{odd}>B$, where $<g_1^{odd}>$ is a constant determined from the slope in Fig 2(c). The even contribution $<\delta^{even}(B_o)>$ appears linear in $V$ (Fig. 2(e)) but notably we do not see a dependence of $<\delta^{even}(B_o)>$ on $B_o$, illustrated in Fig. 2(f) where the slopes, $d(<\delta^{even}(B_o)>)/dV$, are constant with $B_o$. This indicates that $<g_1^{even}(B)>$ is of the form $<g_1^{even}(B)> = <g_1^{even}>$, where $<g_1^{even}>$ is a constant. As the first key result of this Letter, we therefore conclude that the functional dependence of the nonlinear conductance has the form:

$$\langle g(B,V) \rangle = \langle g_0^{even}(B) \rangle + \langle g_1^{even} \rangle V + \langle g_1^{odd} \rangle BV \quad (4)$$

where $<g_0^{even}(B)>$ is the average magnetoconductance in linear response and the coefficients $<g_1^{odd}(B)>$ and $<g_1^{even}(B)>$ can be found through the analysis shown in Fig. 2. The term $<g_0^{odd}(B)>$ is not included in Eq. (4) because reciprocity requires $g_0^{odd}(B) = 0$, consistent with the fact that the measured $<g_0^{odd}(B)> \approx 0.012$ $e^2/h$ is comparable to the noise in our measurements (typically $\sim 1\%$ of $<g_0^{even}(B)>$). Equation (4) is an empirical result which describes the behaviour of the nonlinear conductance for high $B$, $B >> \Phi_o/A$, and high $V$, $V >> E_s/e$.

Having established this basic form, it is now possible to investigate the role of device asymmetry in the nonlinear contributions in Eq. (4). The constants $<g_1^{even}>$ and $<g_1^{odd}>$ were determined for all billiards shown in Table 1. The experimentally measured value $<g_0^{odd}>$ is also included in Table 1 as a baseline for experimental noise. Significantly, devices C and D, which are both UD asymmetric, have significantly larger $<g_1^{odd}>$ values than devices A and B which lack UD asymmetry. This result is consistent with basic symmetry arguments [4,5]; in the case of UD asymmetry, electrons directed by the magnetic field to the top of the billiard encounter a different potential to those directed downward, leading to an asymmetry in the conductance with respect to reversal of $B$. A similar phenomenon occurs for $<g_1^{even}>$, where the only billiard asymmetric with respect to reversal of $V$, device B, has the largest asymmetric contribution to the part of the nonlinear conductance associated with a reversal of bias.

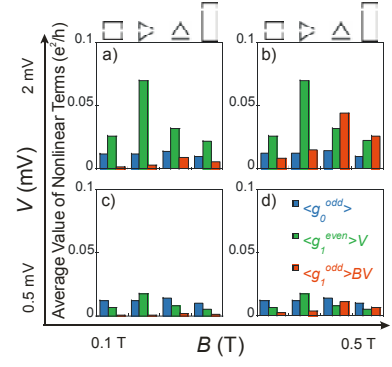

FIG. 3. (Color online) The nonlinear terms $<g_0^{odd}>$ (blue), $<g_1^{even}>V$ (green) and $<g_1^{odd}>BV$ (red) for each of the devices (indicated by the illustrations above the plots) for both low and high $V$ and $B$.

The role of device asymmetry becomes more apparent when the nonlinear terms $<g_1^{even}>V$ and $<g_1^{odd}>BV$ are plotted with $V$ and $B$ for all billiards in Fig. 3. Significant rectification effects are seen for device B with increasing bias (see panels (c) and (a)), while the asymmetric terms for devices C and D grow significantly with $B$ and are maximized for the combined effect of large $B$ and $V$ (panel b). At high $B$ and $V$, significant symmetry breaking is seen for all devices which contain intentional asymmetry and the magnitude of $<g_1^{odd}>BV$ approaches the order of magnitude of the fluctuations. For example, the $<g_1^{odd}>BV$ term for device C is half the size of the fluctuations (0.04 $e^2/h$) at $B = 0.5$ T and $V = 1$ mV. Importantly, note that we observe different values of $<g_1^{odd}>$ depending on the billiard, suggesting that conductor geometry plays a role in nonlinear effects [12].

So far our study has focused on the asymmetry present in the local features of the conductance fluctuations. To further investigate the breakdown of conductance symmetry, we extend our study to the statistical qualities of the asymmetric nonlinear MCF by investigating the spectral density of the fluctuations. For all the billiards presented here, the power spectra of the MCF display a $1/f^{\alpha}$ scaling, where $f = 1/\Delta B$ is the magnetic frequency and $\alpha$ is the spectral exponent of the power law scaling relationship and characterizes the entire spectral content of the fluctuations [13]. In Fig. 4 we show that $\alpha$ increases with $V$. A similar increase in $\alpha$ is seen with temperature for MCF measured at $V = 0$ mV and the dependence of $\alpha$ on $V$ is consistent with an increase in overall electron temperature due to the applied bias [14]. Unexpectedly, however, α evolves asymmetrically with $B$ and $V$ for devices C and D (Fig. 4(c) and 4(d)), indicating that the entire spectral density of the MCF evolves with $B$ and $V$ in an asymmetric manner. This translation of the device asymmetry to such a holistic quality of the MCF is intriguing. Strikingly, we find that α is asymmetric in $B$ and $V$ only for the UD asymmetric billiards (Fig. 4(c) and 4(d)), but not for the LR asymmetric billiards (Fig. 4(a) and 4(b)). One way of interpreting this observation is to consider the geometry of electron injection into the billiard, depending on the direction of $B$ and $V$ (+$B$ defined into the page and +$V$ to the left), as illustrated by the arrows in Fig. 4. For example, in the UD asymmetric device D MCF associated with electrons that initially enter the upper half of the billiard consistently show a smaller α value than MCF associated with electrons that enter the lower half (notice the reversal of the solid and dashed lines at zero bias in Fig. 4(d)). In contrast, for device B, where the upper and lower half and the billiard are nominally the same, no such asymmetry in $B$ is observed (see Fig. 4(b)). In other words, the symmetry, or lack thereof, inherent in the billiard geometry with respect to a reversal of $B$ translates into symmetry, or lack thereof, in the spectral content of the fluctuations. This effect is not captured by the $BV$ term in the nonlinear conductance, and we are unaware of any theory which predicts this behavior.

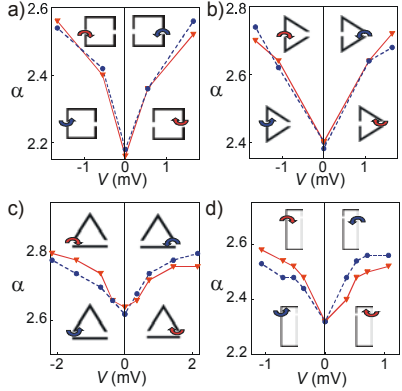

FIG. 4. (Color online) (a) – (d) The spectral exponent α of the MCF power spectrum as a function of ±$V$ for ±$B$ for devices A – D, respectively. α increases with bias for all devices, but unexpectedly α evolves in a manner that is asymmetric in $B$ and $V$ for devices C and D (both are UD asymmetric), while this asymmetry is not notable in the A and B devices (neither is UD asymmetric). The blue circles and red triangles are taken at +$B$ and –$B$ respectively. Curves are provided as a guide to the eye. Also provided are illustrations indicating the injection conditions for each of the four cases, where +$V$ is directed to the left and +$B$ is into the page.

The analysis used here is based on simple symmetry arguments and is thus general and can be used in both the classical and quantum regimes. In order to characterize the asymmetric contribution to the nonlinear conductance up to the high field ($B >> B_c$) and high voltage ($V >> E_s/e$) regime, we averaged the fluctuations (arising from quantum interference effects) in the measured asymmetry with $B$ over several correlation fields. In contrast, the recent theories [6,7] predict the coefficient associated with the nonlinear term proportional to $BV$ for $V < E_s/e$ and in the $B$ field region below one correlation field. In order to determine if these distinct quantities are signatures of the same underlying physical processes, a more comprehensive theory is needed which extends to higher $B$ and $V$.

Note added. During completion of this manuscript we became aware of related experimental work in carbon nanotubes [15] and quantum dots in the $B < B_c$ field range [16].

*Acknowledgments*. We thank M. Büttiker and D. Sánchez for useful discussion. Financially supported by an NSF IGERT (C.A.M.), an NSF CAREER award (H.L.), and a Cottrell scholarship (R.P.T).
*Corresponding author: cmarlow@uoregon.edu

device size to phase-breaking length for D is more than double that of C leading to increased dephasing. In linear response the average size of the fluctuations for D is less than half the size of the fluctuations for C, consistent with an increase in decoherence effects. Smaller fluctuations would lead to overall smaller observed asymmetries due to nonlinear contributions.

13   $1/f^{\alpha}$ scaling is a signature of fractal behaviour, see for example, J.-F. Gouyet, *Physics and Fractal Structures* (Springer-Verlag, New York, 1996). The fractal behaviour of the MCF observed in these devices is consistent with previously observed fractal conductance fluctuations in similar systems [10,12,17].

14   H. Linke et al., Phys. Stat. Sol., 318 (1997); M. Switkes et al., Applied Physics Letters **72**, 471 (1998); A. G. Huibers et al., Physical Review Letters **83**, 5090–5093 (1999); C. A. Marlow et al., To be published in Springer Proceedings in Physics Series as proceedings of HCIS-14, Chicago, July 24-29 (2005).
15   J. Wei et al., cond-mat/0506275 (2005).
16   D. M. Zumbuhl et al., cond-mat/0508766 (2005).
17   C. A. Marlow et al., submitted to Phys. Rev. B (2005).
18   J. P. Bird et al., Surface Science **361/362**, 730 (1996).